\documentclass[twocolumn,amsmath,amssymb]{revtex4}

\usepackage{graphicx}
\usepackage{dcolumn}
\usepackage{bm}

\begin{document}

\title{Large capacitance enhancement and negative compressibility of two-dimensional electronic systems at LaAlO$_3$/SrTiO$_3$ interfaces}
\author{Lu Li$^1$, C. Richter$^2$, S. Paetel$^2$, T. Kopp$^2$, J. Mannhart$^2$, R. C. Ashoori$^1$
}
\affiliation{
$^1$ Department of Physics, Massachusetts Institute of Technology, Cambridge, Massachusetts 02139, USA.\\
$^2$ Center for Electronic Correlations and Magnetism, University of Augsburg, Augsburg, 86135, Germany
}

\date{\today}
\begin{abstract}
Novel electronic systems forming at oxide interfaces comprise a class of new materials with a wide array of potential applications. A high mobility electron system forms at the LaAlO$_3$/SrTiO$_3$ interface and, strikingly, both superconducts and displays indications of hysteretic magnetoresistance\cite{HwangNature2004,HwangNature2002,MannhartScience2006,ScienceSC2007,NatMaterHystMR2007}. An essential step for device applications is establishing the ability to vary the electronic conductivity of the electron system by means of a gate.  We have fabricated metallic top gates above a conductive interface to vary the electron density at the interface. By monitoring capacitance and electric field penetration, we are able to tune the charge carrier density and establish that we can completely deplete the metallic interface with small voltages. Moreover, at low carrier densities, the capacitance is significantly enhanced beyond the geometric capacitance for the structure. In the same low density region, the metallic interface overscreens an external electric field. We attribute these observations to a negative compressibility of the electronic system at the interface. Similar phenomena have been observed previously in semiconducting two-dimensional electronic systems \cite{PRLEisenstein1992,PRBEisenstein1994,PRBSiKravchenko,PRLShapira96,PRLJiang2000,PRLAllison2006,SSCAshoori1992}. The observed compressibility result is consistent with the interface containing a system of mobile electrons in two dimensions \cite{PRLMillis2006, MRSMannhart2008}.

\end{abstract}


\maketitle                   

The origin of the conductivity of in the LAO/STO system is still under debate (see, e.g.,  \cite{PRLMillis2006,PRLPickett2009,MRSMannhart2008}). Hard x-ray photoelectron spectroscopy measurements, scanning conductance microscopy experiments, and delta-doping studies demonstrate that the conductivity occurs within a few nanometers from the interface \cite{PRLHardXray2009,NatMaterSCM2008,PRLDeltaDoping2009}. One may expect that at the interface the carrier density can be tuned by the electric field effect, similar to Si-MOSFET and GaAs/AlGaAs-heterostructures. Indeed, prior experiments have demonstrated that the conductivity and the superconducting transition temperature can be changed by applying electric fields from a metallic gate on the back of the SrTiO$_3$ substrate \cite{MannhartScience2006,NatureFieldEffect2008}. However, a recent study on the resistance and Hall effect suggests that back-gating primarily alters the effective disorder, rather than changing the carrier density \cite{HwangDisorderFieldEffect2009}. Thus, it is important to demonstrate the tuning of the carrier density by electric fields applied across the LaAlO$_3$ cap layer,  rather than the SrTiO$_3$ substrate. Determination of the transport characteristics of the top-gate-tunable oxide interface will shed light on the ground state of the electron system at the oxide interface and lead to significant possible applications \cite{JAPCapacitanceKopp}.

\begin{figure}[t]
\includegraphics[width=6.5in]{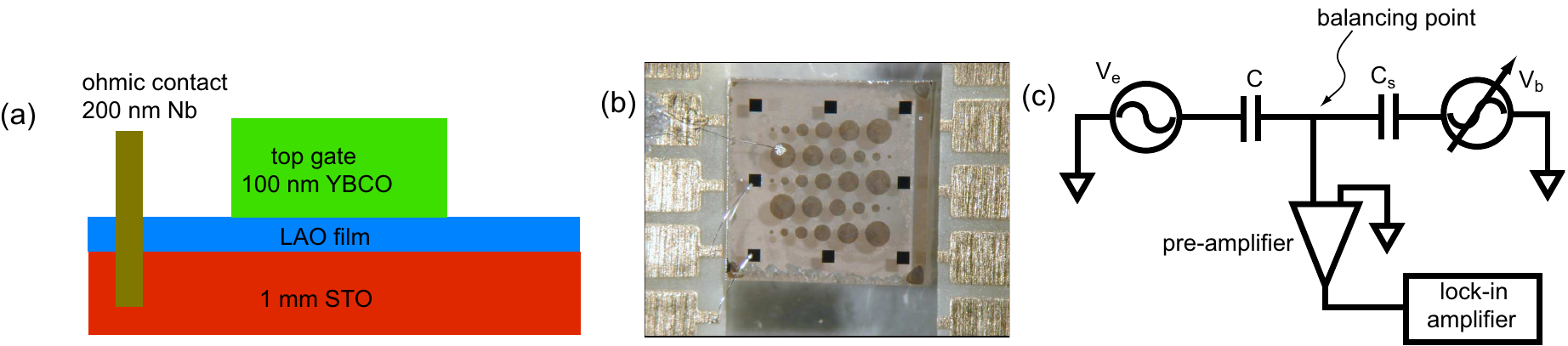}
\caption{\label{setup}(color online) Sample layout and the capacitance bridge setup. (Panel a) Sketch of the oxide interface layout. A thin layer of LaAlO$_3$ ( 10 u.c. or 12 u.c. thick ) is deposited onto the top of a SrTiO$_3$ substrate. YBa$_2$Cu$_3$O$_{7-\delta}$ (YBCO) top gates are deposited and patterned above the LaAlO$_3$ layer, and Nb ohmic contacts are deposited close to the corners of the wafer. (Panel b) Picture of a YBCO/LaAlO$_3$/SrTiO$_3$ sample with leads attached. The wafer is square with side-length 5 mm. The diameter of the YBCO circular top gates varies between 50 $\mu$m and 500 $\mu$m. (Panel c) Setup sketch of the capacitance bridge. In one arm of the bridge,  $C$ stands for the sample capacitor, excited by an AC excitation voltage $V_e$. In the other arm, another AC voltage $V_s$ with the same frequency is applied to a standard capacitor $C_s$. The signal at the balancing point is measured with a pre-amplifier and a lock-in amplifier. During the measurement, with the phase and the amplitude of the excitation voltage $V_e$ stable, the balancing voltage $V_s$ is varied both in phase and in amplitude to null the signals at the balancing point.
}
\end{figure}

We fabricated capacitor devices on LAO/STO heterostructures by growing in-situ YBa$_2$Cu$_3$O$_{7-\delta}$ (YBCO) films on the surface of the LAO. As shown in Fig. \ref{setup}(a), we then patterned the YBCO films on top of the LAO surface, and Nb films were deposited into ion-milled holes to make ohmic contacts to the conductive layer at the LAO/STO interface. Technical details of
the sample fabrication are described in the Methods section. Each top gate forms a two-terminal capacitor. The thickness of the LAO layer is either 10 or 12 unit cells (u.c.). Fig. \ref{setup}(b) displays a photograph of a sample, with electrical leads wire-bonded to the ohmic contacts near the edge of the wafer and a lead connected to a circular top gate.

The two-terminal capacitance of the capacitor device is measured with a home-built capacitance bridge. The bridge automatically adjusts exciting voltages on the two arms of the bridge to null the charge signal at the balancing point, enabling us to measure the capacitance in the frequency range between 1 Hz to 2 MHz, with $\sim$~20 microradian resolution in the phase measurement of the impedance.  We can also vary the DC voltage on the top gate and track how the capacitance varies with the gate voltage (see Methods).

\begin{figure}[ht]
\includegraphics[width=6.5in]{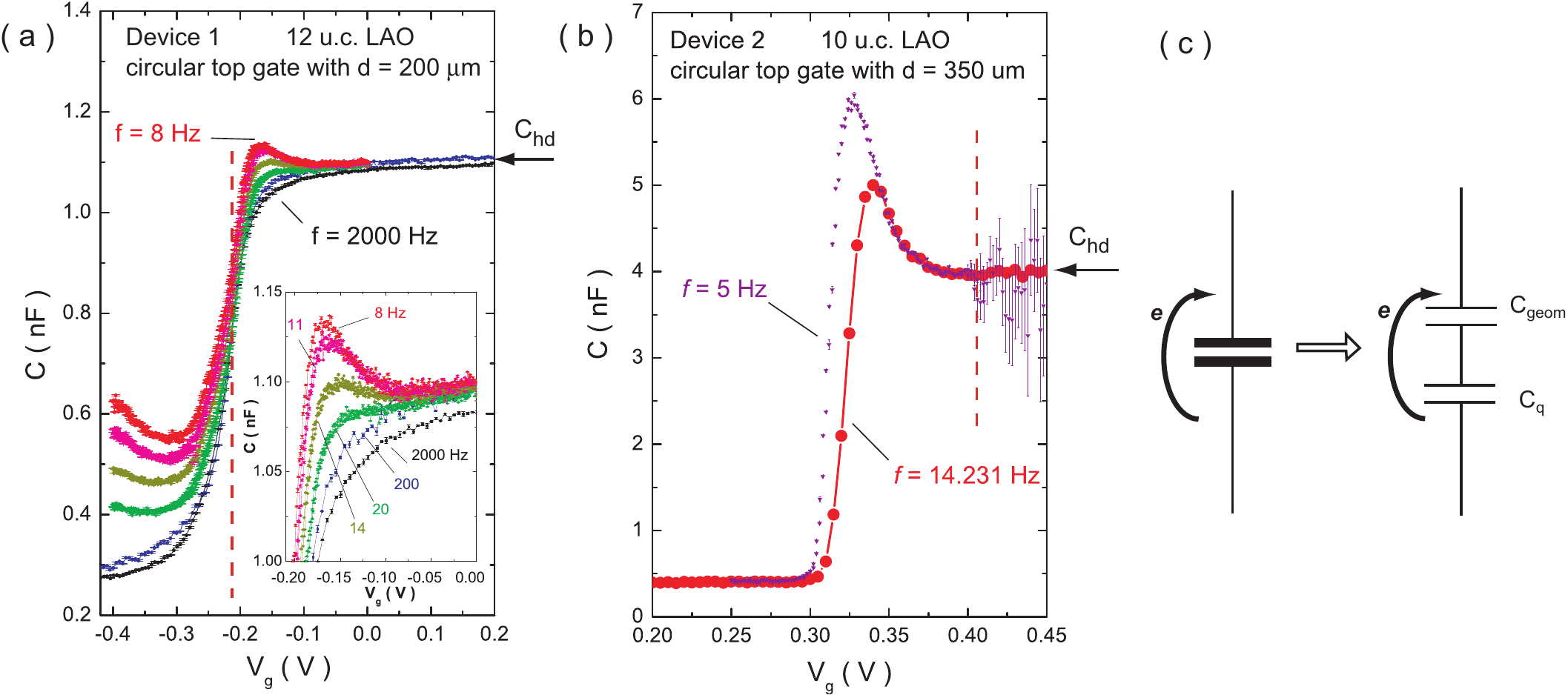}
\caption{\label{cV}(color online) The gate voltage dependence of the capacitance.  (Panel a) The $C-V_g$ curve of Device 1 with diameter 200 $\mu$m and 12 u.c. thick LaAlO$_3$ layer. The curves are taken at $T = 4.2$ K at various frequencies between 8 Hz and 2000 Hz. The capacitance at high carrier density $C_{hd}$ is marked with an arrow. As shown in the inset for the region near the depletion, the capacitance values at low carrier densities increase and exceed $C_{hd}$. At $V_g$ smaller than - 0.22 V, marked by the red dashed line, we note that the capacitance curves are frequency dependent, which is probably a result of DC leakage through the LaAlO$_3$ barrier. (Panel b) The $C-V_g$ curve of Device 2 with diameter 350 $\mu$m and 10 u.c. thick LaAlO$_3$. The capacitance at high carrier density $C_{hd}$ is again marked with an arrow. The enhancement of the capacitance is even more pronounced as the peak of $C$ is about 40\% larger than $C_{hd}$ at $f$ = 5 Hz. For this device, a leakage current develops at higher $V_g$ and the 5 Hz capacitance measurement becomes noisier. The red dashed line indicates a leakage resistance of 5 M$\Omega$. (Panel c) Equivalent circuit of geometric capacitance $C_{geom}$ and the quantum capacitance $C_q$. The charging of the electrodes is represented by arrows. The geometric capacitance $C_{geom}$ is determined by the dimensions of our device, marked as $C_{hd}$ in Panel a and b. The quantum capacitance $C_q$ is defined as $Ae^2 \frac{dn}{d\mu}$, where $A$ is the effective area of the capacitor, $n$ is the carrier density, and $\mu$ is the chemical potential. The quantum capacitance reflects how the chemical potential $\mu$ changes as $n$ varies. It is determined by the kinetic energy, the exchange energy, and the correlation energy of the electronic system \cite{JAPCapacitanceKopp}.
}
\end{figure}

We performed capacitance $C$ vs. top gate voltage $V_g$ measurements at $T = 4.2$ K, and typical characteristics are shown in Fig. \ref{cV}. Device 1 is fabricated on a sample of 12 u.c. LAO. On the top surface of the LAO film a circular top gate with a diameter of 200 $\mu$m is patterned. We measured the capacitance between the top gate and the conductive interface at frequency $f$ ranging from 8 Hz to 2000 Hz while varying $V_g$. Fig. \ref{cV}(a) displays the $C$ vs. $V_g$ curves of Device 1. Similar curves for Device 2 are shown in Fig.~\ref{cV}(b), where the LAO thickness is 10 u.c., and the diameter of the top gate is 350 $\mu$m. Substantial leakage (resistances of less than 5 M$\Omega$) through the LAO barrier occurs at voltages to the left of the red line drawn in Fig. \ref{cV}(a) and to the right of the red line drawn in Fig. \ref{cV}(b). Data in these regions were not used in further analysis because the leakage resistance drops to a magnitude similar to the in-plane resistance of the device. The leakage current alters the voltage across the device and changes the charge distribution within the sample. By examining the capacitance at high electron densities at which interaction effects on the capacitance are usually small, we determine the dielectric constant of the insulating LAO layer in the device.  Given the capacitance values at $V_g = 0$  and at large positive $V_g$ for Device 1, the size of the gate, and the thickness of the LAO film of each device,  we determine the dielectric constant $\epsilon \sim$ 18 $\epsilon_0$, where $\epsilon_0$ is the dielectric constant of vacuum. While this value is lower than the dielectric constant of bulk single crystals of LAO, it agrees with measurements of thin films grown by MBE \cite{APLLAOMBE}.

Both of the capacitor devices show a sharp depletion signal at certain gate fields. For Device 1, at $V_g < -0.2$ V, the capacitance is strongly diminished, suggesting that the electron density underneath the circular gate is substantially suppressed by the electric field. In earlier works \cite{NatureFieldEffect2008,HwangDisorderFieldEffect2009}, the capacitance values between the conductive interface and the back gate on the STO were tracked as the back gate voltages $V_b$ were changed. The $C-V_b$ curves did not show clear drops at the metal-insulator transition. This fact was used to argue against the claim of field-effect \cite{HwangDisorderFieldEffect2009}, but it may be attributed to incipient ferroelectricity and charge trapping in the STO \cite{NatureFieldEffect2008}. In contrast, by applying voltages at top gates across the LAO layers, we do not involve the STO substrate since we keep both the interface and the back gate DC-grounded. In this case, we observe a clear and sharp depletion in the capacitance curves, which suggests the electron carrier density at the interface is tuned to zero with the applied gate electric field.

At low temperature the interface underneath the gate in Device 2 (10 u.c. thick LAO layer) is not conductive at zero gate voltage. A positive $V_g \geq 0.28$ V is needed to create a conducting channel underneath the top gate, even though the regions of the sample away from the YBCO circular pads are still conductive (the two-terminal resistance between the Nb ohmic contacts is of the order of 500 $\Omega$ at 4.2 K). In testing these samples during at least four different coolings from room-temperature to 4.2 K for each sample, we noticed that the depletion voltage may vary slightly with thermal cycling. However, the depletion voltage of devices with 12 u.c. LAO is always negative, whereas that for devices with 10 u.c. LAO layers it is positive.  This fact suggests that the YBCO top gate tends to deplete the interface underneath the gate. The farther the interface is away from the gate, the smaller the depletion effect. This behavior might be attributed to the difference between the work function of YBCO and the vacuum, to the polar nature of the YBCO ionic layers, or to the strain effect due to the lattice mismatching.

Close to depletion we observe a large enhancement of the capacitance, which is shown in detail in the inset of Fig. \ref{cV}(a). At its peak, the capacitance is considerably larger than $C_{hd}$, the capacitance at high densities. $C_{hd}$ $\sim$ $C_{geom} = \epsilon\frac{A}{d}$, where $\epsilon$ is the electric permittivity, $A$ and $d$ are the cross section and the distance between of electrode of the capacitor. (The difference between $C_{hd}$ and $C_{geom}$ is usually small and is typically determined mainly by the effective mass. Below, all the $C_{geom}$ values are treated to be same as $C_{hd}$.) Such an enhancement of capacitance is known to appear in high-mobility 2D-electron-system (2DES) and 2D-hole-system (2DHS) samples. However, the size of the enhancement is usually a few percent \cite{PRBSiKravchenko,PRLEisenstein1992,PRLShapira96,PRLAllison2006}. In these systems, mobile carriers confined in 2D over-screen the external electric field and enhance the capacitance at low carrier densities. Charging a capacitor $C$ with charge $e$ does not simply require a change in voltage across the capacitor $\delta V = {e}/{C_{geom}}$. Because of the finite ``thermodynamic density of states'' (TDOS) \cite{PRBTDOSLee82}, $\frac{dn}{d\mu}$ in the 2D system, charging the capacitor requires an extra voltage $\frac{d\mu/dn}{eA}$, where $\mu$ is the chemical potential, and $n$ is the carrier density. Thus, as shown in Fig. \ref{cV}(c), the measured capacitance $C$ is
the in-series combination of the geometric capacitance $C_{geom}$ and the quantum capacitance
$C_q = Ae^2 \frac{dn}{d\mu}$:

\begin{equation}
\frac{1}{C} =  \frac{1}{C_{geom}}+\frac{1}{C_q}
 					  =  \frac{1}{C_{geom}}+\frac{1}{Ae^2 \frac{dn}{d\mu}}.
\label{Capacitance}
\end{equation}
Therefore, with precise knowledge of the geometric capacitance, the measurement of the capacitance provides a sensitive probe to determine the TDOS. This quantity has been widely studied in 2D semiconductor devices \cite{PRBSiKravchenko,PRLEisenstein1992,PRLShapira96,PRLJiang2000,PRLAllison2006,SSCAshoori1992}.

For positive $\frac{dn}{d\mu}$, the quantum capacitance diminishes the measured capacitance from its classical (geometric) value.  The observed enhancement of $C$ instead reveals that $\frac{dn}{d\mu}$ is negative rather than positive. This phenomenon is known as ``negative compressibility'' \cite{PRLEisenstein1992}. In two dimensions,  the compressibility is given by $\frac{1}{A}\frac{dA}{dS} = n^{-2}\frac{dn}{d\mu}$, where $S$ is the line pressure on the perimeter of an area $A$. (This definition of the compressibility does not include direct electron repulsion from the 2D layer.) In semiconductor electronic systems, negative compressibility has been found to result largely from the effects of the exchange interaction which diminishes the energy required to add an electron to a capacitor plate \cite{PRBEisenstein1994}.  In general, other effects may also lead to a measured negative electronic compressibility: the correlation energy or the shifting of the center position of electrons in finite-width quantum wells due to the application of the gate voltage, and the interaction of the charge carriers with non-electronic degrees of freedom such as phonons \cite{PRBEisenstein1994, JAPCapacitanceKopp}. In a free-electron system, the correlation enhancement is calculated to be smaller than the effect of the exchange interaction \cite{PRB2DEGCeperley}, and the effect of carrier position shift should be small in a system where mobile carriers are confined to only a few atomic layers.

The capacitance upturn at small $n$ was observed at low frequency ($f < 15$ Hz). As shown in the inset of Fig. \ref{cV}(a), as the frequency $f$ decreases from 20 kHz to 20 Hz, the capacitance is independent of frequency over a large range of $V_g$. However, at low densities, the peak in $C$ continues to grow larger and moves to lower gate voltage as the frequency is reduced. This frequency dependence is reproduced in other devices. Fig.~\ref{cV}(b) shows the $C-V_g$ curves of Device 2 with 10 u.c. LAO film. For measurements at two different frequencies, the upturns coincide between 0.33 V $\leq V_g \leq$0.4 V. Below $V_g$ = 0.33 V, the curve taken at $f$ = 14.231 Hz diminishes, whereas that at $f$ = 5 Hz keeps increasing and drops at lower $V_g$. The overlapping part of these two capacitance curves is the DC limit of the capacitance curve. In this regime, the out-of-phase charging signal stays small and displays a completely different trend (see the supplement). Measuring at frequencies as low as 2 Hz, we have not found a limit to the divergence.  For these low frequencies, strikingly, we observe a greater than 40\% enhancement in the sample capacitance, larger than has ever been previously observed, even in high mobility GaAs-based devices \cite{PRBSiKravchenko,PRLJiang2000,PRLAllison2006}. Presumably, at sufficiently low frequencies, disorder will limit the sharpness of the upturn (as it does in GaAs samples). Finally, the sheet resistivity of the layer at the capacitance peak at 5 Hz exceeds 10 M$\Omega$ per square.

\begin{figure}[ht]
\includegraphics[width=6.5in]{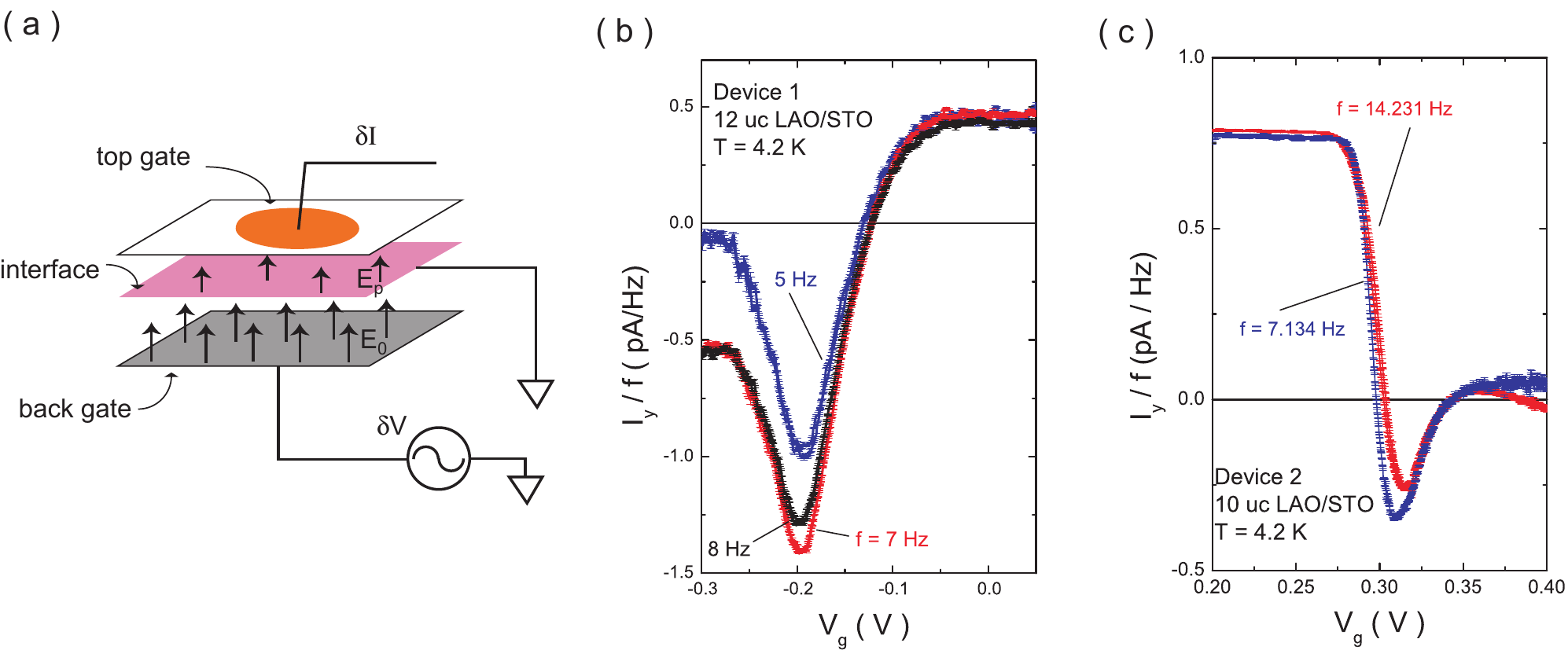}
\caption{\label{penetration}(color online) Penetration field measurement. (Panel a) Sketch of the penetration electric field setup. With the interface grounded, we apply an external electric field $E_0$ from the back gate to the interface and detect the electric field that penetrates through the grounded interface layer. The penetration field $E_p$ is determined by measuring the current from the top gate of the device. In principle, the penetration current $I_y$ should be proportional to the frequency $f$, when $f$ is small enough to allow full charging of the interfacial layer.  (Panel b) The penetration current $I_y$ divided by the measurement frequency for Device 1 at $T = 4.2$ K measured at three excitation frequencies. The LaAlO$_3$ thickness is 12 u.c. At $V_g$ near zero, the penetration current is proportional to $f$, and is constant over a broad range -0.05 V $\leq V_g \leq$ 0.05 V. For $V_g < $ -0.18 V, $I_y / f$ displays a frequency dependence, likely due to effects of current leakage through the LAO layer. (Panel c) The penetration current $I_y$ divided by the measurement frequency for Device 2 at $T = 4.2$ K measured at two selected frequencies. The LaAlO$_3$ thickness is 10 u.c.  At 0.31 V $< V_g <$ 0.34 V, the negative values of the penetration current demonstrate that the penetration field $E_p$ points in the opposite direction to the external field $E_0$; the conductive interface overscreens the external fields at low carrier densities. This overscreening comes from electron-electron interaction, and it is generally called ``negative compressibility''. At larger $V_g >$ 0.37 V , we notice again a $f$-dependent signal which develops at gate voltages where there is measurable current leakage through the LAO barrier.
}
\end{figure}

An interesting consequence of the negative quantum capacitance is that the interface overscreens the external electric fields. To explore the possible existence of this overscreening effect, we carried out the field penetration measurement using a technique developed by Eisenstein \cite{PRLEisenstein1992,PRBEisenstein1994}. As shown in Fig. \ref{penetration}(a), the interface was grounded, and we measured the current response on the top gate as we applied a small AC excitation on the back gate. The current is proportional to the electric field penetrating from the back gate through the grounded interface (see Methods). The results are shown in Fig. \ref{penetration}(b) and (c), in which $\frac{I_y}{f}$ is proportional to $\frac{d\mu}{dn}$, the inverse of the compressibility. A negative divergence of the penetration field happens close to depletion, where the carrier density is below $2\times10^{12}$ cm$^{-2}$. Here the carrier density is estimated by integrating the capacitance between the top gate and the interface at the lowest measurement frequency. The negative current observed at low carrier densities reveals that the electric field penetrating through the metallic interface is negative, which arises from overscreening of the external field by the mobile electrons at the LAO/STO interface. To further explore the electronic properties of the interface, we follow the analysis of Ref. \cite{PRLJiang2000}, which has shown that, for sufficiently low frequency $f$, the penetration current $I_y$ goes as

\begin{equation}
I_y = \frac{2\pi fC_1C_2V_{ac}}{C_q},
\label{IyCq}
\end{equation}
where $C_1$ is the geometric capacitance between the back gate and the interface normalized by the area of the top gate, $V_{ac}$ is the excitation voltage on the back gate, and $C_2$ is the geometric capacitance between the top gate and the interface. We measure $C_1$ and $C_2$ directly. For Device~1, $C_1 =$ 4.67 pF, $C_2 = 1.1$ nF, and $V_{ac}=$ 20 mV. For Device~2, $C_1 =$ 14.3 pF, $C_2 =$ 4.01 nF, and $V_{ac}=$10 mV.

For Device 1, in the gate region $V_g <$ -0.18 V, the penetration current $I_y$ is no longer proportional to $f$. This might be attributed to the DC leakage in this region, although we do not know the precise mechanism by which a DC leakage would affect the measurement. The same deviation occurs with Device~2 in the positive gate region at $V_g > $ 0.37 V. Consequently, we exclude these regions in the analysis below.

\begin{figure}[bh]
\includegraphics[width=6.5in]{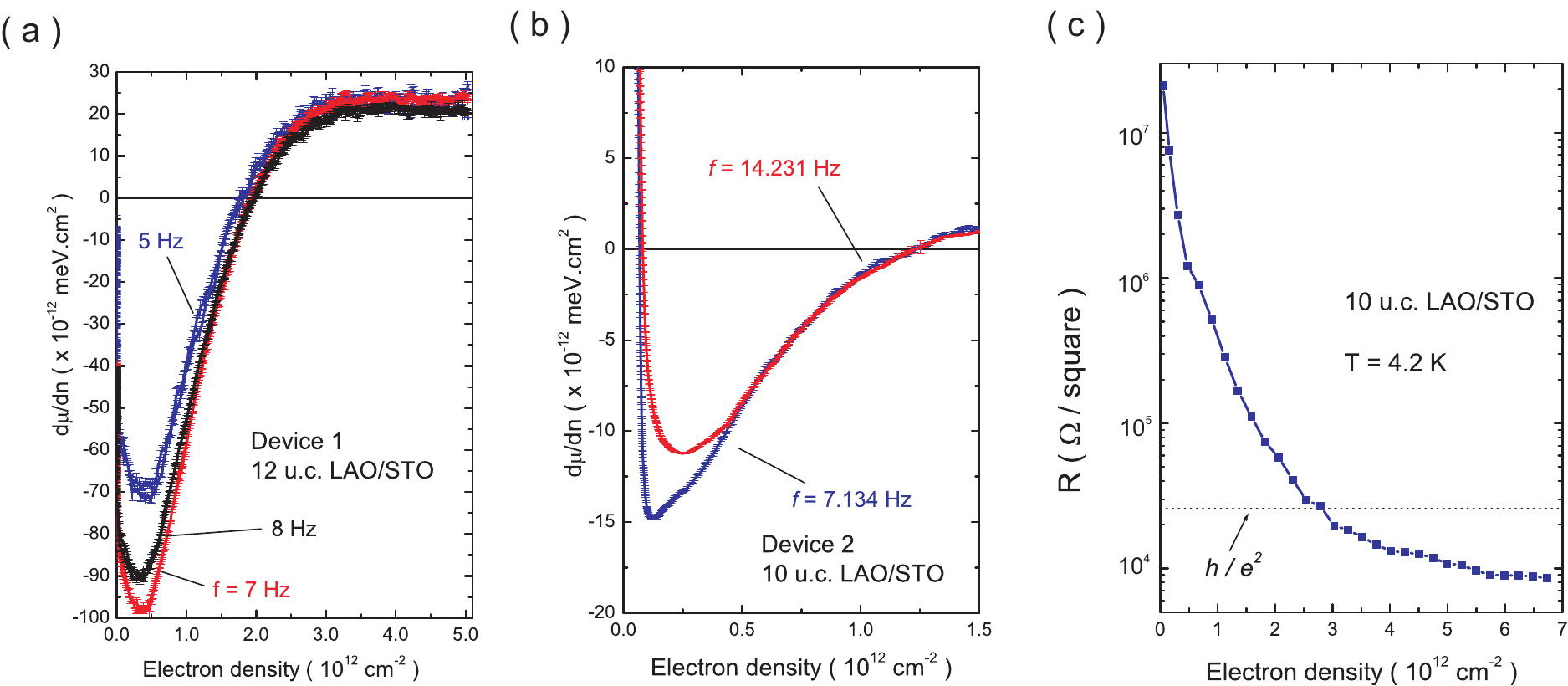}
\caption{\label{chemicalpotential}(color online) The inverse thermodynamic density of states $\frac{d\mu}{dn}$ determined from penetration field measurements on Device 1 (Panel a) and Device 2 ( Panel b). Using the measured capacitance $C$ between the top gate and the oxide-interface layer, the electron density at the interface is determined by integrating the $C$ vs. $V_g$ curve at the lowest frequency we achieved. Note for the electron density $n$ below 2 $\times$ 10$^{11}$ cm$^{-2}$, there is uncertainty in determining $n$, since we cannot measure the capacitance at sufficiently low frequencies for full charging of the layer. (Panel c) The lateral resistivity vs. the electron density $n$ of a different device fabricated in the same 10 u.c. LaAlO$_3$/SrTiO$_3$ wafer. The diameter of this device is 500 $\mu$m. The resistivity is determined by tracking the frequency dependence of the out-of-phase charging signals. ( See the supplement )
}
\end{figure}

We determine the quantum capacitance $C_q$ and use Eq. \ref{Capacitance} to compute $\frac{d\mu}{dn}$. Fig. \ref{chemicalpotential}(a) and (b) show the density dependence of $\frac{d\mu}{dn}$ of Device 1 and Device 2, respectively. The zero of density is also somewhat uncertain, since the measurement frequency at low densities is not sufficiently low to fully charge the device. To lowest order in the electron-electron interaction the ground state energy consists of the average kinetic energy $\frac{\hbar^2\pi}{2m^*}n^2$ and the exchange energy $-\frac{1}{3}\sqrt{\frac{2}{\pi}}\frac{e^2}{\pi\epsilon}n^{\frac{3}{2}}$.
Thus, for a homogeneous 2D electron gas in a very narrow quantum well, the inverse compressibility is given by (see, for example, \cite{PRB2DEGCeperley, JTEPBello81})
\begin{equation}
\frac{d\mu}{dn} = \frac{d^2E}{dn^2}=\frac{\hbar^2\pi}{m^*}-\sqrt{\frac{2}{\pi}}\frac{e^2}{4\pi\epsilon}n^{-\frac{1}{2}},
\label{compressibility}
\end{equation}
where $n$ is the density of the mobile electrons, $m^*$ is their effective mass, and $\epsilon$ is the relative dielectric constant of the electron system. At electronic densities for which $\frac{d\mu}{dn}$ approaches the regime of negative values, correlations may dominate this expression. Their effects on the compressibility are described in Refs.~\cite{PRB2DEGCeperley,JAPCapacitanceKopp} for homogenous electron systems.

Examination of the data in Fig. \ref{chemicalpotential}(a) (Device 1) reveals a broad region of nearly constant $\frac{d\mu}{dn}$, which we interpret as the constant expected from Eq. \ref{compressibility} for the regime of high electron densities. Under this assumption, the effective mass $m^*$ is found to be 0.12$\pm$ 0.02 $m_e$. This estimated value is an order of magnitude lower than the theoretical prediction, in which the effective mass is determined by the coupling of the Ti$^{4+}$ orbitals and is found to be about 1.6 -- 4 $m_e$ \cite{condmatSTM2DELMannhart,InfraredLAOSTO}. The cause of this large disagreement has not been identified. In our experimental setup, any offset in current reading due to the stray capacitance may affect the estimation of $m^*$. However, we measured this offset and found that it is less than 25\% of the measured $I_y$ signal for Device 1 (see the supplement), which could lead to 20\% underestimation of $m^*$. One possible cause for the small measured $m^*$ might be that the YBCO top gate adjacent to the LAO layer affects the electronic state at the interface underneath the gate, for example, due to the strain induced by mismatching lattices. Further, we emphasize that Eq.~\ref{compressibility} holds for a homogeneous electron gas. For inhomogeneous electron systems additional terms arise in the energy functional and the capacitance \cite{JAPCapacitanceKopp} which represent the interaction between the charge carriers and all other degrees of freedom such as local potentials or phonons. In oxides, due to the strong confinement of the electron system at the interface \cite{condmatSTM2DELMannhart}, these terms are expected to be enhanced as compared to standard, large bandwidth semiconductors. Yet, because the terms cannot be reliably quantified at present, they have not been included in this analysis, but can affect the estimate of the effective mass.

Attempts to fit to $\frac{d\mu}{dn}$ data with Eq. \ref{compressibility} at lower densities do not yield a satisfactory fit.  As the density is lowered, the onset of negative compressibility is sharper than predicted by Eq. \ref{compressibility}. We note that in this low density range, the average distance between electrons equals at least the thickness of the LAO layer. Thus, image charges distort the standard $\frac{1}{r}$ dependence of the Coulomb interaction, thereby affecting the $n$-dependence of the compressibility \cite{CapacitanceShklovskii}. Moreover, as shown in Fig. \ref{chemicalpotential}(c), near depletion the sheet resistivity of the interfacial layer becomes much larger than $h/e^2$. Therefore, the electrons are localized at low densities and the system may be outside of the valid range of the Hartree-Fock model. Strong disorder usually broadens the charging threshold for the structure and blurs away any sign of negative compressibility, as observed in GaAs heterostructures \cite{PRLAllison2006}. The observation of negative compressibility in our disordered oxide interface devices is therefore surprising.

The observation of negative compressibility suggests that the interface behaves as a quantum mechanically two-dimensional electron system, with electrons having no degree of freedom to move in the direction perpendicular to the interface. If there were several uncoupled layers in the conductance channel, each layer would successively screen the penetration field, and the field penetration would be much smaller than that measured (Fig. \ref{penetration}). If these layers are coupled, one would expect strong dipolar screening of externally applied fields and, contrary to our observations, we should see no penetration field in the high density limit.

The large capacitance enhancement that results from the negative compressibility offers a possible gating mechanism to switch transistors using small gate voltages. This could diminish heating issues in future logic devices \cite{JAPCapacitanceKopp,CapacitanceShklovskii,Solomon}. We emphasize that the observed large enhancement of capacitance $C$ is also affected by the large geometric capacitance $C_{geom}$. Taking derivatives on both sides of Eq. \ref{Capacitance}, we get $\frac{\delta C}{C} = C\frac{\delta C_q}{C_q^2}$. As a result, the device with larger $C_{geom}$ per unit area is expected to have a larger relative enhancement of the total capacitance. Therefore, an interface capacitor device with a thinner LaAlO$_3$ layer could lead to even larger capacitance enhancement at low carrier densities. As material quality improves it is plausible that at low densities the effective gate-layer capacitance could increase to several times the geometric value. Transistors fashioned from the oxide interface could then have an effectively very ``high-$\kappa$'' electrode, allowing creation of small transistors that switch at low voltages and minimal gate to channel leakage \cite{Solomon}.

{\it\large{Acknowledgment}}

The authors gratefully acknowledge helpful discussions and interactions with O. E. Dial, M. M. Fogler, R. Jany, P. A. Lee, L. S. Levitov, A. J. Millis, B. Shklovskii, P. Solomon, and J.-M. Triscone. This work was supported by ARO-54173PH, by the National Science Foundation through the NSEC program, by the DFG (TRR 80) and the EC (OxIDes), and by the Nanoscale Research Initiative.  L. Li would like to thank the support of the MIT Pappalardo Fellowships in Physics.

{\it\large{Methods}}

The LaAlO$_3$/SrTiO$_3$ heterostructures were grown at the Augsburg University using pulsed laser deposition with in-situ monitoring of the LaAlO$_3$ layer thickness by reflection high energy electron diffraction. The single crystalline SrTiO$_3$ substrates were TiO$_2$ terminated. Their typical lateral size is  5 x 5 mm$^2$ and their thickness is 1 mm. The LaAlO$_3$ layers were grown at an oxygen pressure of 8x10$^{-5}$ mbar at 780 $^\mathrm{o}$C to a thickness of 10 unit cells or 12 unit cells. For the top gate, we deposited an in-situ 100 nm YBa$_2$Cu$_3$O$_{7-\delta}$ (YBCO) layer at 760$^\mathrm{o}$C and etched it to circular shape with H$_3$PO$_4$ acid. After deposition of the YBCO, the samples were cooled to room temperature in 0.5 bar of oxygen.  The sputtered ohmic Nb contacts filled holes patterned by etching with an Ar ion-beam.

The capacitance of our devices is measured at MIT with a home-built capacitance bridge. For a two-terminal capacitor device $C$, the top gate is electrically connected to one plate of a standard capacitor $C_s$ ( typically 10 pF), and the voltage or current signal is monitored at this ``balance point'' of the bridge, marked in Fig. \ref{setup}(c). A 1 mV AC excitation voltage is applied to the interface of the device, the amplitude and phase of a balancing voltage on $C_s$ are adjusted to minimize the AC voltage signal on the balance point. With the phase accuracy better than 20 microradian, we can measure the capacitance signals with 20 ppm resolution. The capacitance $C$ is inferred from the balancing voltage. A DC voltage $V_g$ is applied to the top gate of the device to tune the carrier density of the interface underneath the gate.

The field penetration effect is measured following the experimental setup of Ref. \cite{PRBEisenstein1994} and \cite{PRLJiang2000}. As the interface is grounded through an ohmic contact, we apply an AC excitation voltage to the back of the SrTiO$_3$ substrate, and monitor the AC current coming out of the top gate using a Ithaco 1211 preamplifier or the preamplifier of the SR830 lock-in amplifier. The capacitance channel $I_y$ is proportional to the electric field penetrating through the interface \cite{PRLEisenstein1992}.

\clearpage

\end{document}


\newcommand{\CFF}{\ensuremath{^4\mathrm{CF}}}
\newcommand{\CFT}{\ensuremath{^2\mathrm{CF}}}
\newcommand{\CFO}{\ensuremath{^1\mathrm{C}}}
\newlength{\scolwid}
\newlength{\dcolwid}
\setlength{\scolwid}{3.3in} 
\setlength{\dcolwid}{6.5in} 

%
%
\textbf{\Large Supplementary Material}
\small
\setlength{\baselineskip}{3.76mm}
\doublespacing
\section{Out-of-phase signals in the capacitance measurements}

In the capacitance measurements, both the in-phase charging signal and the out-of-phase charging signal provide useful information. As shown in the electronic setup sketch in Figure 1 in the main text, the in-phase balancing measures the sample device capacitance $C$. Sometimes a small out-of-phase component is needed to null the bridge, due to the conductance or the inductance of the sample device. This out-of-phase component is particularly important in finite frequency region when the sample is close to depletion. In this case, the lateral resistance is so large that the capacitor device does not charge fully in response to the ac voltage at the measurement frequency of the capacitance bridge. A simple model of this situation is a two-element circuit with a capacitor $C$ and a resistor $R$ in series. In charging the capacitor $C$ through the resistor $R$, the in-phase capacitance signal drops to half at frequency $f_R = \frac{1}{2\pi RC}$, and the out-of-phase signal peaks at the same frequency.

\begin{figure}
\ifthenelse{\boolean{figs}}{{\resizebox{6.5in}{!}{\includegraphics{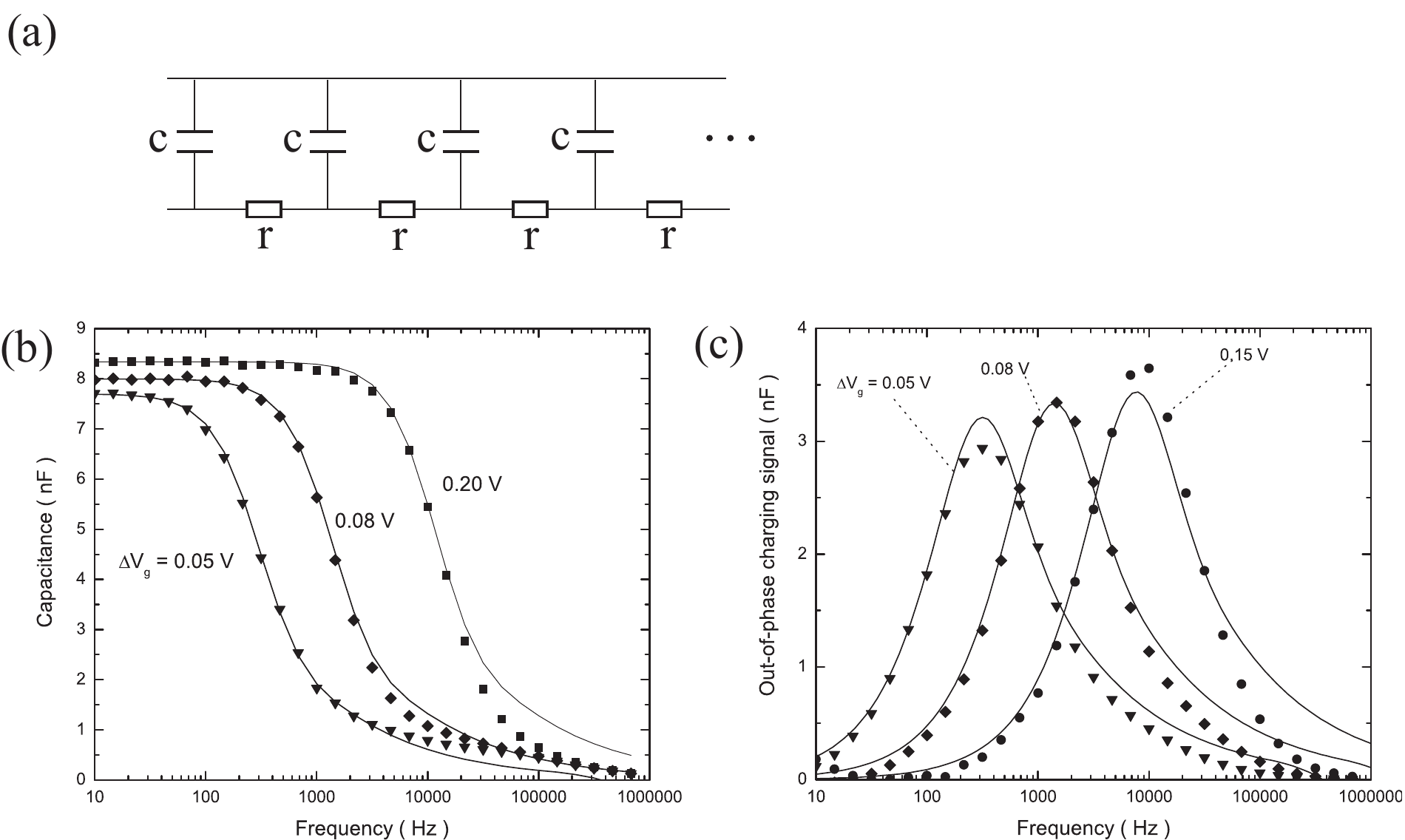}}}}{}
\caption{\textbf{Charging capacitor devices in finite frequency.}
(Panel a) Sketch of the charging model of a distributed RC network. (Panel b) The in-phase $C$ vs. $f$ curves taken at selected gate voltages and $T$ = 4.2 K. The thickness of LAO is 10 u.c. and the diameter of the circular top gate is 500 $\mu$m. The smooth curves are results of fitting with the distributed RC network. ( Panel c ) The frequency $f$ dependence of the out-of-phase charging signal taken in the same conditions. The results of fitting using the distributed RC network are also plotted in solid lines.
 }
\label{cf}
\end{figure}

The charging of our devices is more complicated than a simple RC circuit, because of the finite cross-section of our circular device and the large value of the lateral resistance. Our device is similar to a distributed RC network, shown in Figure \ref{cf}(a), where the capacitances per unit area are in parallel with each other and connected by a finite lateral resistance $r$. At the edge of the circular device, the network is connected to the rest of the conductive interface, finally to the ohmic contacts. The charging response of this kind of distributed RC network was calculated in detail in Ref. \cite{PRBGoodall} for circular disk, which finds that the peak in the out of phase signal occurs at $f_R = 1.3 /CR_l$, where $C$ is the capacitance in DC limit, and $R_l$ is the lateral resistivity per square.

The distributed RC network model provides a better fit to our measurements  than the simple RC model, especially for frequencies beyond the peak in the out of phase signal. An example of the frequency dependence of the in-phase and out-of-phase capacitance charging signals are shown in Figure \ref{cf}(b) and \ref{cf}(c) for a device in the 10 u.c. LaAlO$_3$/SrTiO$_3$ sample. The diameter of this circular device is 500 $\mu$m. A top gate voltage $V_g$ is varied to tune the carrier density $n$, which is determined by the geometric capacitance $C_{geom}$ and $\Delta V_g$, the difference between the current gate voltage and the completely depletion gate voltage. For clarity, only three sets of curves are plotted in Figure \ref{cf}(b) and \ref{cf}(c). At low $f$, $C$ stays constant and the out-of-phase signal is close to zero. As $f$ increases above the loss peak frequency $f_R$, $C$ rapidly diminishes, whereas the out-of-phase signal peaks at the same frequency. The fits of the in-phase $C - f$ curves are using the distributed RC network are plotted in solid lines. As shown in Figure \ref{cf}(b), the measured $C - f$ curves display a long tail at $f > f_R$. The measured drop-off of $C$ at $f_R$ is less abrupt than that given by the simple RC model, whereas the distributed RC network model tracks closely of the diminishing at $f$ up to $f_R$. In the frequency region higher than $f_R$, the measured $C - f$ curves deviates from the prediction of the network model.

The same trend is also seen in the out-of-phase signal. In Figure \ref{cf}(c), the out-of-phase charging signals are compared with the modeling curves using the RC network model. For each data set at certain carrier density, between 10 Hz and $f_R$, all of the modeling curves track the measured results. But the measured higher frequency part is smaller than the prediction. The reason might be attributed to the giant dielectric constant of the SrTiO$_3$ substrate. The electronic system at the STO/LAO interface is the conductance channel connecting the capacitor device to the ohmic contact. The shunt capacitance between the interface and the grounded back gate is quite large ( $\sim$ 3.7 nF, see the second part of the supplement material) due to giant dielectric constant of SrTiO$_3$ at low temperature. The lateral two-point resistance is in the order of 100 - 1000 $\Omega$. As a result, there is a considerable loss of the signal and phase-rotation in high frequency region.

\begin{figure}
\ifthenelse{\boolean{figs}}{{\resizebox{3in}{!}{\includegraphics{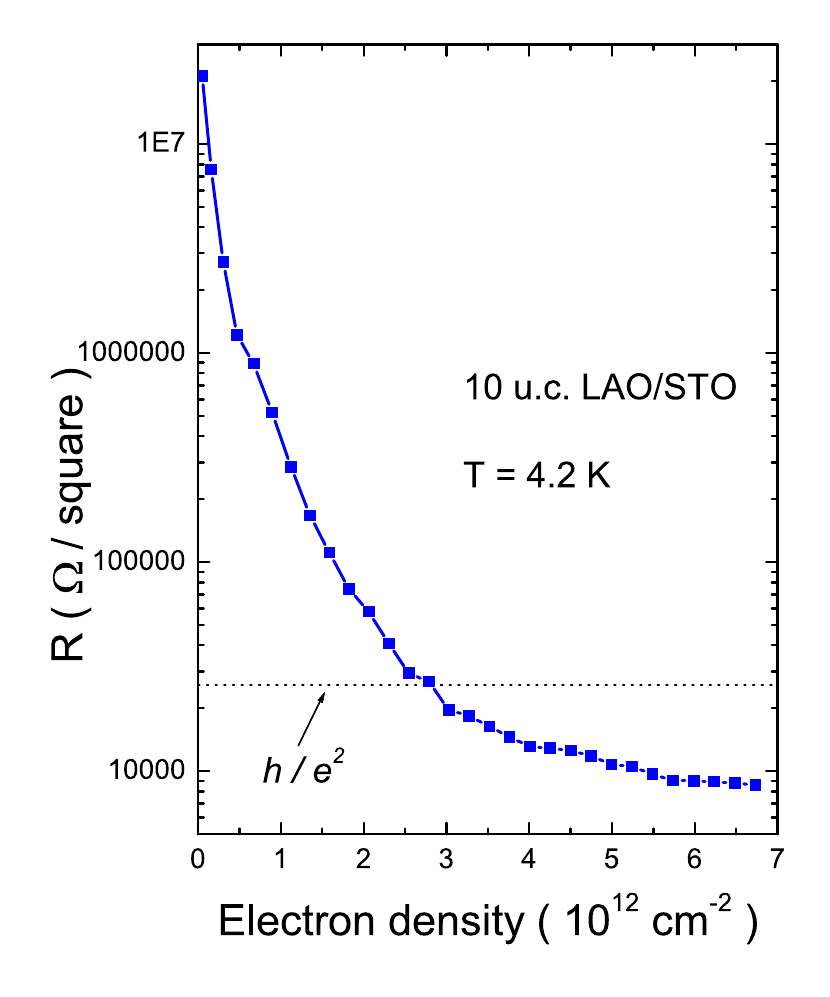}}}}{}
\caption{\textbf{Lateral resistance of the interface in the capacitor devices.}
Lateral resistance is inferred from the frequency dependence curves of the in-phase and out-of-phase charging signals  using the distributed RC network model. The $h/e^2$ value is marked in the dashed line. The electron density is determined by integrating the $C$ vs. $V_g$ curve of the device. The device starts to leak at about $3.5\times10^{12}$ cm$^{-2}$. However, for the density region shown here, the leakage distorts the out-of-phase signal only at $f < $100 Hz, which does not affect the determination of $f_R$ or $R_l$.
 }
\label{Rn}
\end{figure}

Although the fitting is not perfect at high frequencies, we can still obtain a good measurement of $f_R$ from fitting to the lower frequency portions of each of the data sets. The lateral resistance is determined qualitatively as $R_l = 1.3 / Cf_R$ with the distributed RC network model. The result is plotted in Figure \ref{Rn} against the electron density $n$. The lateral resistance $R_s$ increases from about 10 $k\Omega$ per square to about 20 $M\Omega$ per square near the depletion. We note that at density $n$ lower than 2 -3 $\times$ 10$^{12}$ cm$^{-2}$, the lateral resistance is already bigger than the quantity $h/e^2$.

\section{Capacitance measurements using the back-gate and back-gating}
\begin{figure}
\ifthenelse{\boolean{figs}}{{\resizebox{7.5in}{!}{\includegraphics{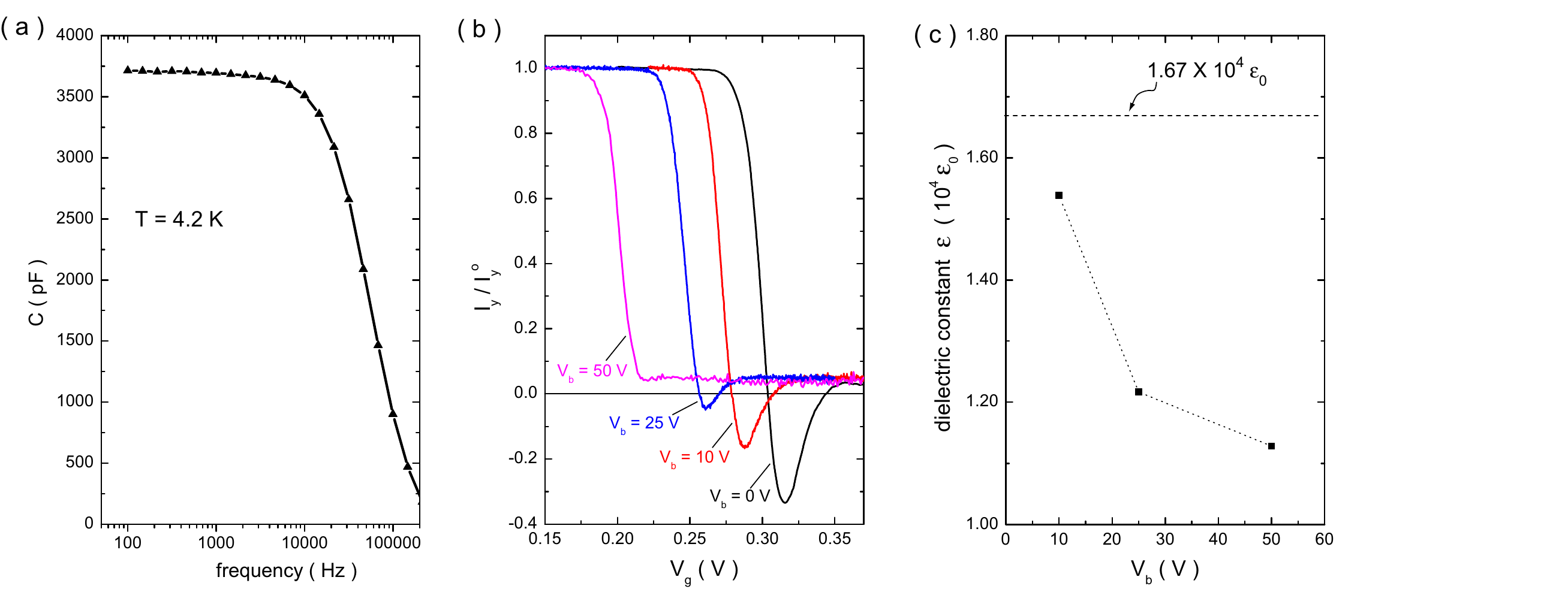}}}}{}
\caption{\textbf{The effect of back gate.}
(Panel a) The frequency dependence of the capacitance between the back gate and the interface taken at $T$ = 4.2 K. Using the low frequency capacitance value, we calculate that the dielectric constant of the SrTiO$_3$ substrate is about 1.67$\times$10$^4\epsilon_0$. (Panel b)  Normalized penetration currents $I_y / I_y^0$ vs. top gate voltage $V_g$ on Device 2 taken at different back gate voltages $V_b$. The temperature is 4.2 K and the excitation voltage is 10 mV. The normalized factor $I_y^0$ is the penetration current when the interface is completely depleted by the top gate. The applied negative back gate $V_b$ changes the depletion voltage at the top gate. (Panel c) Using the capacitance value from the top gate to the interface, we infer that a 0.1 Volt change corresponds to a carrier density change of $2.67\times10^{12}$ cm$^{-2}$. Using this density change, we infer that the dielectric constant of the SrTiO$_3$ substrate at various back gate voltage $V_b$ (solid square). The zero bias dielectric constant is shown as the dashed line. As $V_b$ increases, the dielectric constant of the substrate decreases greatly from the zero bias value.
 }
\label{backgate}
\end{figure}

We carried out capacitance measurements to determine the dielectric constant of the SrTiO$_3$ substrates in our samples. Figure \ref{backgate}(a) displays the frequency dependence of the capacitance between the back gate and the interface. At low frequency, the capacitance $C$ stays almost constant. In the high frequency end, $C$ curve decreases because of the finite conductance of the leads and the interface. The low frequency capacitance value $C \sim$ 3700 pF is used to estimate the dielectric constant of the SrTiO$_3$ substrate. Given the size of the substrate (5 mm $\times$ 5 mm  and 1 mm thick), the dielectric constant of the substrate is about 1.67$\times$10$^4\epsilon_0$, where $\epsilon_0$ is the vacuum permittivity.

The dielectric constant of the SrTiO$_3$ substrate is also measured by applying electric field from the back gate to tune the carrier density at the interface. Figure \ref{backgate}(b) shows the effect of the back gate on the penetration field measurements. We applied different negative DC voltages $V_b$ on the back gate to bring more charges into the interface and tracked the penetration current vs. the top gate $V_g$ traces. The depletion top-gate voltage changes from 0.3 V at zero back gate voltage $V_b$, to about 0.25 V at $V_b =$ -25 V, then to about 0.20 V at $V_b =$ -50 V. Thus, the carrier density indeed changes when the back gate voltage is applied. Since the geometric capacitance does not change at different back gate voltage $V_b$, we can infer the change of the carrier density as the depletion voltage $V_g$ varies. Given the capacitance between the top gate and the interface and  the size of the top gate in Device 2,  the 0.1 V change in $V_g$ corresponds to a change in the carrier density 2.67$\times$10$^{12}$cm$^{-2}$. This yields a dielectric constant of the SrTiO$_3$ substrate at different back gate voltage, which is displayed in Figure  \ref{backgate}(c). The values at finite $V_b$ are smaller than the dielectric constant ( 1.67$\times$10$^4\epsilon_0$ ) obtained by direct capacitance reading at zero bias. The difference reflects the nonlinear dielectric response of SrTiO$_3$ substrate \cite{NatureFieldEffect2008,HwangDisorderFieldEffect2009}. The larger the electric field on the substrate is, the smaller polarization it generates. Thus the dielectric constant is smaller at larger applied voltage.

We note that a side-effect of applying back gate voltage $V_b$ is the reduction of the negative divergence of the penetration current $I_y$. It might be a result of the disorder brought by the back-gating, as observed in Ref \cite{HwangDisorderFieldEffect2009}. Further studies are needed to understand the effect of the back-gating.

\section{Stray background in the penetration field measurement}

\begin{figure}
\ifthenelse{\boolean{figs}}{{\resizebox{3in}{!}{\includegraphics{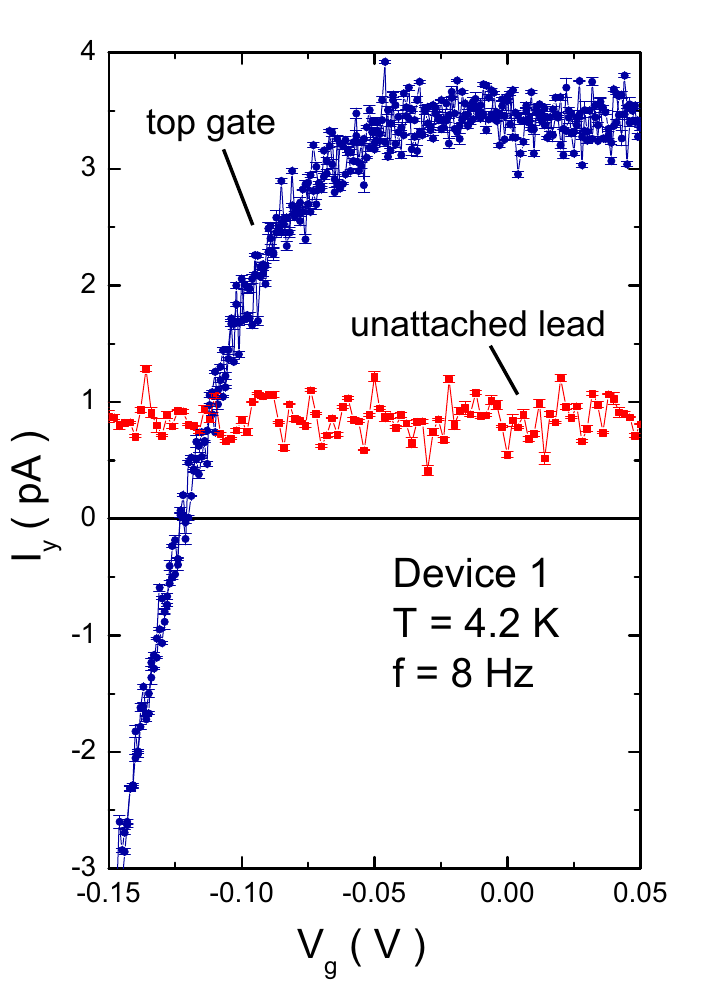}}}}{}
\caption{\textbf{The stray capacitance signal in the penetration current measurements.}
 The current signal between a back gate and an unattached lead is shown for comparison to the penetration current $I_y$ measured from the top gate for Device 1. The temperature $T$ = 4.2 K and the excitation voltage $V_{ac}$ is 20 mV.
 }
\label{straybackground}
\end{figure}

An offset in the penetration field measurements affects our estimation of the effective mass of the interface. The offset is caused by the stray capacitance between electric leads. We worked to reduce the stray capacitance by using shielded cables up to very close to the sample. As a check of the effectiveness of this shielding, we measured the stray capacitance from an unattached electric lead close to a measured pad. The capacitance reading is in the order of 10$^{-3}$ pF, whereas a typical capacitance value of our device is in the order of 1 nF.  For the penetration field configuration, the measured current to the unattached lead is plotted in Figure \ref{straybackground} and compared with the penetration current data of Device 1 as the voltage varies.  The measurement frequency is 8 Hz and the temperature is 4.2 K. As shown in Figure \ref{straybackground}, the stray capacitance only gives a small positive background, which is less than 25 \% of the penetration current signal of Device 1. The background signal does not change as the gate voltage is applied to this unattached wire. Note for free electrons the penetration current $I_y$ is proportional to $\frac{d\mu}{dn}$, which equals $\frac{\hbar^2\pi}{m^*}-\sqrt{\frac{2}{\pi}}\frac{e^2}{4\pi\epsilon}n^{-\frac{1}{2}}$ if we neglect the effect of correlations. The offset of $I_y$ can shift the magnitude of $\frac{d\mu}{dn}$ equally at different $V_g$, but does not affect the curvature of the curve of $\frac{d\mu}{dn}$ vs. $n$. Therefore, the stray capacitance background may make us underestimate the effective mass by at most 20 \%.

\newpage

\newpage
\newpage

\newpage

\newpage

